\documentclass{PoS}

\title{Long-term TeV Observations of the Gamma-ray Binary HESS J0632+057 with VERITAS}

\ShortTitle{Observations of HESS J00632+057 with VERITAS}

\author{\speaker{G.Maier for the VERITAS Collaboration}\thanks{veritas.sao.arizona.edu}\\
        Deutsches Elektronen-Synchrotron (DESY)\\
        E-mail: \email{gernot.maier@desy.de}}

\abstract{
The gamma-ray binary HESS J0632+057 has been observed at very-high energies for almost ten years by all major systems of imaging atmospheric Cherenkov telescopes.
We present new observations taken by the VERITAS observatory and an updated analysis of the previously published observations.
HESS J0632+057 has been detected by VERITAS with a total significance $>20\sigma$ at energies above 350 GeV in about 200 hours of observations.
The exposure covers now almost the entire orbital period of  315 days including the flux enhancements at phases $\approx$0.35 and, for the first time, detected with VERITAS,  at phases $\approx$0.75. 
The results are discussed along with simultaneous observations by the X-ray satellite {\em Swift} XRT.}

\FullConference{The 34th International Cosmic Ray Conference,\\
		30 July- 6 August, 2015\\
		The Hague, The Netherlands}

\begin{document}

\section{Introduction}

Gamma-ray binaries are a small subclass of binary systems identified by variable emission in the GeV-to-TeV range, and a peak in their spectral energy distribution at gamma-ray energies (see \cite{Dubus-2013} for a review).
The presence of high-energy photons shows that they are efficient particle accelerators. 
Where and how gamma rays are produced in these systems is not known at this point and a number of possibilities are discussed in the literature.
For accretion-powered sources, the acceleration sites might be connected to internal shocks in the jet or the jet-stellar wind interaction zone.
Pulsar-powered scenarios propose the production of shocks by the interaction of a relativistic pulsar wind with the stellar wind or the circumstellar disk.
The presence of a strong stellar wind, wind inhomogeneities ('clumping'), a dense circumstellar disk, the pulsar wind, and the orbital movement results in dynamical changes of the physical conditions with a large number of free parameters which complicates the interpretation of the existing data.
More observations at high energies are clearly necessary to get a better understanding of the underlying processes.
We present in these proceedings new observations of the gamma-ray binary  HESS J0632+057 by the VERITAS observatory in 2013-2015 and updated results from observations during the period  2006-2012.

HESS J0532+057 is a gamma-ray binary consisting of a massive star of type B0 Vpe  (MWC 148=HD 259440) and a compact object (neutron or black hole) located at a distance of 1.1-1.7 kpc \cite{Aragona-2010}.
It is a point-like very-high energy ($>$100 GeV) gamma-ray emitter serendipitously discovered by the High Energy Stereoscopic System (H.E.S.S.) during observations of the Monoceros Loop supernova remnant in 2004 and 2005 \cite{Aharonian-2007}.
Gamma-ray observations by VERITAS revealed evidence of variability \cite{Acciaro-2009}, while 
long-term X-ray observations using the Swift X-ray Telescope (XRT) displayed periodical flux modulations \cite{Bongiorno-2011}. 
These X-ray observations firmly established the binary nature of the object.
Further gamma-ray observations by VERITAS, H.E.S.S.~\cite{Aliu-2014} and MAGIC \cite{Aleksic-2012} revealed a  pattern of variability, with significant gamma-ray emission above 1 TeV in two different phases ranges of the orbit.
The orbital period of the binary system of $(315\pm5)$ days has been derived from X-ray data \cite{Aliu-2014}, the orbital solution obtained from radial spectroscopy measurements at optical wavelengths points towards an eccentric orbit ($e\approx0.83$, \cite{Casares-2012}). 
HESS J0632+057 is the only gamma-ray binary which has not been detected at MeV-GeV energies with the Fermi LAT \cite{Caliandro-2013}.


\section{Gamma-ray and X-ray Observations}

VERITAS is a gamma-ray observatory sensitive to photons in the energy range from 85 GeV to $>$30 TeV.
It is located at the Fred Lawrence Whipple Observatory in southern Arizona (31  40 N, 110 57 W) and consists of an array of four imaging atmospheric-Cherenkov telescopes
(for details on the instrument and its performance see \cite{Staszak:2015, Park:2015, WWW}).
HESS J0632+057 has been observed by the VERITAS observatory for a total of 200 hours between 2006 December and 2015 January at energies above $\approx$200 GeV.
The results presented here have been obtained with updated analysis algorithms compared to \cite{Aliu-2014}, with the most important change being the application of a boosted-decision tree based gamma-hadron separation algorithm to the data.
The sensitivity improvement in comparison to the previous analysis based on box cuts is roughly 15-20\% \cite{Park:2015}.

At X-ray energies, {\em Swift}-XRT monitored HESS J0632+057 at 0.3-10 keV from 2009 January to 2015 January.
The observations have typical durations of $\approx$4-5 ks taken at intervals between one week and several months. 
The data has been analysed using standard {\em Swift} tools \cite{Swift}.

\section{Results}

\begin{figure}[htbp]
\begin{center}
\centering\includegraphics[width=0.85\linewidth]{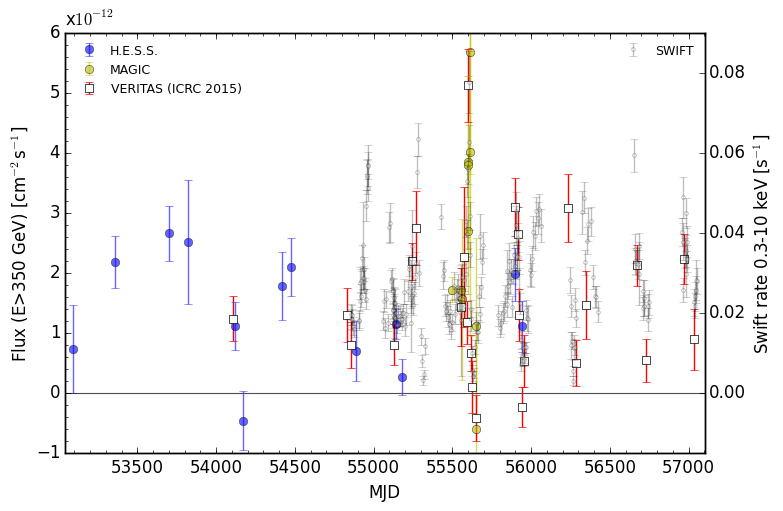}
\caption{Long-term gamma-ray observations of HESS J0632+057 with VERITAS, H.E.S.S. \cite{Aliu-2014}, and MAGIC \cite{Aleksic-2012} at energies $>$350 GeV
and X-ray observations with {\em Swift}-XRT (0.3-10 keV; gray markers).
\label{Fig:longterm}}
\end{center}
\end{figure}

\begin{figure}[htbp]
\begin{center}
\centering\includegraphics[width=0.45\linewidth]{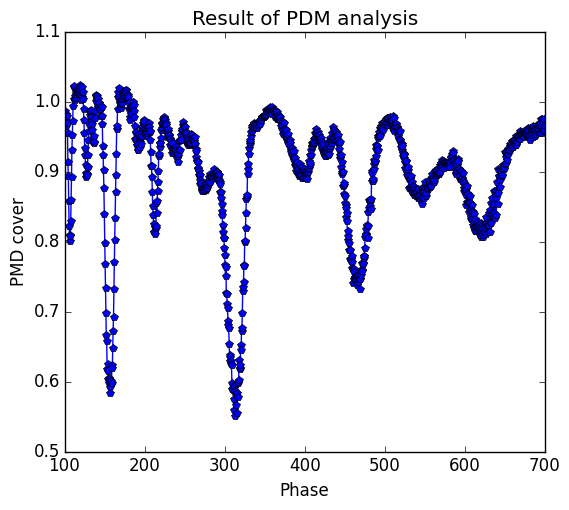}
\centering\includegraphics[width=0.45\linewidth]{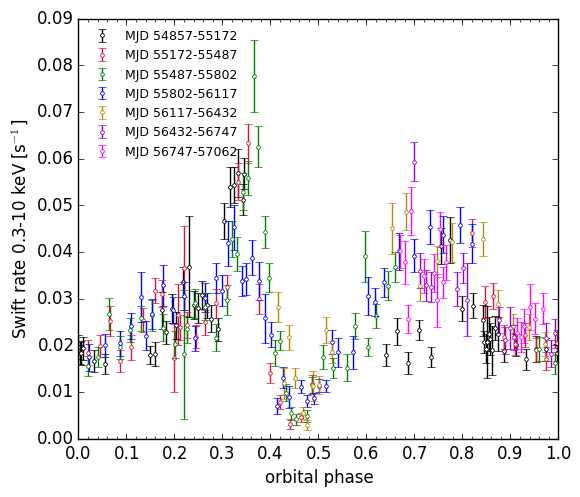}
\caption{Left: Phase Dispersion Minimization (PDM) test statistic vs orbital period (in units of days) calculated from the {\em Swift}-XRT data.
 The minimum value of the test statistic is at 313 days.
Right: Phase folded {\em Swift}-XRT X-ray light curve (0.3-10 keV). 
Different colours indicate different orbital periods.
}
\label{Fig:XRayAnalysis}
\end{center}
\end{figure}

HESS J0632+057 has been detected as a source of gamma rays with a detection significance of $20.5\sigma$ for the complete VERITAS set of observations.
The gamma-ray emission is variable, as can be seen from the observations spanning an interval of almost 10 years (Figure \ref{Fig:longterm} and Table \ref{table:results}).
Periods of flux levels below the detection limit of the instrument alternate with high flux states reaching flux levels of $\approx$3\% of the flux of the Crab Nebula at the same threshold energy of 350 GeV.

The updated set of {\em Swift}-XRT observations allows to reevaluate the determination of the orbital period of the binary.
The variation of the X-ray rate with time is non-sinusoidal, we therefore choose for the analysis the Z-transformed discrete autocorrelation function (Z-DCF; \cite{Alexander-1997}) and, for cross checks, the phase dispersion minimisation method (PDM; \cite{Stellingwerf-1978}; see Figure \ref{Fig:XRayAnalysis} left)  for the analysis.
Both methods give comparable results consistent with the value of $\tau_{orbit}^{2014} = 315^{+6}_{-4}$ days  reported in \cite{Aliu-2014}: 
$\tau_{orbit}^{ZDCF} = 299^{+13}_{-15}$ days and $\tau_{orbit}^{PDM} = 313$ days.
We therefore continue to assume in the following an orbital period of 315 days.
Figure \ref{Fig:XRayAnalysis} (right) shows the phase-folded X-ray light curve, revealing a regular pattern with two emission maxima at phases $\approx$0.35 and  $\approx$0.75, and a marked dip close to apastron at phases $\approx$0.45 (using the orbital solution of \cite{Casares-2012}).
Note the orbit-to-orbit variability of up to a factor of two in X-ray fluxes during the high-state phases. 

\begin{figure}[htbp]
\begin{center}
\centering\includegraphics[width=0.85\linewidth]{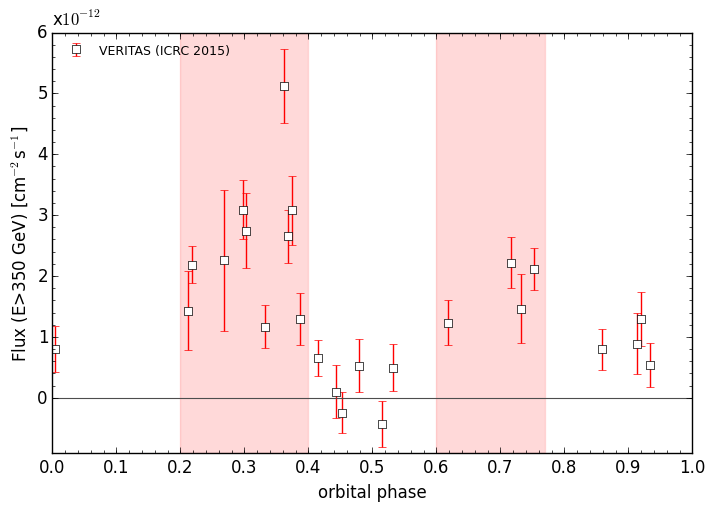}
\caption{Phase-folded VERITAS gamma-ray light curve for energies $>$ 350 GeV using
an orbital period of 315 days.
The shaded areas indicated the selection of phase ranges used for the spectral analysis (see Table 1 and Figure 4).
}
\label{Fig:GRayAnalysis}
\end{center}
\end{figure}

\begin{figure}[htbp]
\begin{center}
\centering\includegraphics[width=0.44\linewidth]{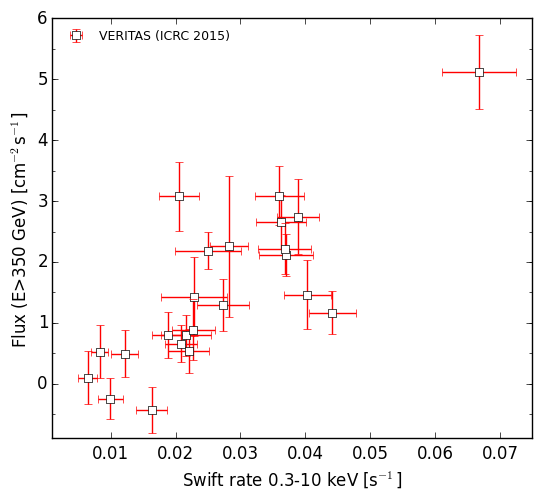}
\centering\includegraphics[width=0.47\linewidth]{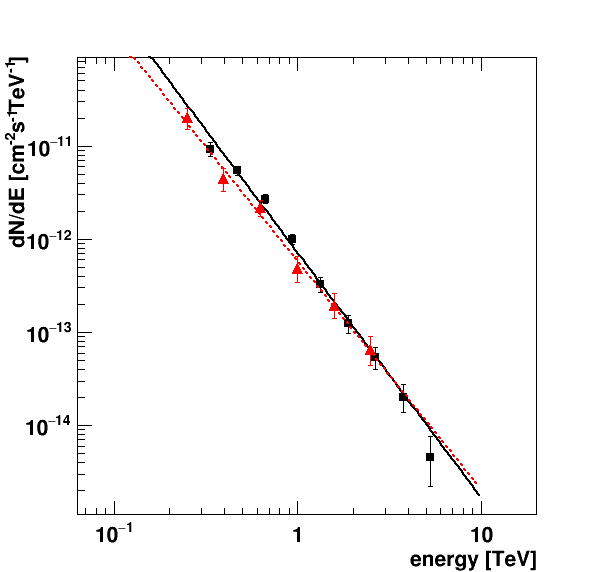}
\caption{Left: Gamma-ray ($>$350 GeV) fluxes vs. X-ray (0.3-10 keV) rates for contemporaneous observations 
(defined here as $\pm2.5$ day intervals around the date of the gamma-ray observations).
Right: Differential energy spectra for gamma-ray photons for the orbital phase ranges 0.2-0.4 (black markers and black line) and 0.6-0.8 (red markers and red line).
The lines show results from fits assuming power-law distributions. Fit results can be found in Table 1.
}
\label{Fig:spectra}
\end{center}
\end{figure}

The phase-folded VERITAS gamma-ray light curve for energies > 350 GeV (Figure \ref{Fig:GRayAnalysis}) shows a variability pattern across the orbit  similar to the X-ray light curve.
The first maximum at phases 0.2-0.4 is brighter than the emission at phases 0.6-0.75, but like the X-ray measurement exhibits  flux variability at similar phases.
The binary is not only clearly detected around the second emission maximum (phases 0.6-0.75 with 9.7$\sigma$ significance), but also at later phases (7.1$\sigma$ significance at phases 0.8-0.2), indicating the possibility of constant low-level emission in the phases ranges 0.8-0.2. 
No significant gamma-ray emission has been detected during the dip in X-ray fluxes around apastron. 
The differential energy spectra in gamma rays during the two maxima can be described by power-law distributions (Figure \ref{Fig:spectra}).
The parameters of the spectral fits are consistent with each other, which points towards similar physical conditions at the gamma-ray emission sites during the high states before and after the apastron phase.

\begin{table}[htdp]

\begin{center}
\begin{tabular}{l|c|c|c|c|c}
{\bf Orbital phase} & {\bf all phases} & {\bf 0.2-0.4}  & {\bf 0.4-0.6} & {\bf 0.6-0.75} &{\bf  0.75-0.2} \\
\hline
{\bf Observation time} (h) & 201.6 & 74 & 46 & 29 & 52 \\
{\bf Significance} ($\sigma$) & 20.5 & 19.2 & 2.5 & 9.7 & 7.1 \\
\hline 
{\bf Flux Normalization} $\Phi_0$  &  & & & & \\
{\bf at 1 TeV} & $4.1\pm0.2$ & $7.18\pm0.43$ & - & $5.58\pm0.74$ & $3.23\pm0.49$ \\
{\bf Photon index} $\gamma$ & $2.69\pm0.06$ & $2.63\pm0.07$  & - & $2.48\pm0.16$  & $2.68\pm0.2$ \\
$\chi^2$/N & 20.9/9 & 13.0/7 & - & 3.6/6 & 3.1/6 \\
\end{tabular}
\end{center}
\caption{
\label{table:results}
Analysis results for energies $>$350 GeV for the phase-folded VERITAS measurements.
The lower three lines of the table show the fit results assuming a power-law distribution $dN/dE = \Phi_0 \cdot E^{-\gamma}$ of the data,
see Fig. 4.
Errors are $1\sigma$ statistical errors.
The flux normalisation constant  $\Phi_0$ is in units of $10^{-13}$ cm$^{-2}$ s$^{-1}$ TeV$^{-1}$.
}
\end{table}

The indicated correlation between X-ray and gamma-ray emission is shown in Figure \ref{Fig:spectra}.
A correlation analysis using the ZDCF method results in a significant correlation ($ZDCF/ZDCF_{\mathrm{error}}=6.8$).
The time lag between X-ray and gamma-ray data  is with $+3.2^{+5.5}_{-7.8}$ days consistent with zero.

\section{Conclusions}

The X-ray and gamma-ray light curve of HESS J0632+057 stands out among the five known gamma-ray binaries.
It shows two maxima, with a sharp minimum in the emission pattern close to the apastron phase.
X-ray and gamma-ray fluxes are correlated. 
This is predicted by simple one-zone leptonic emission models, where 
relativistic electrons lose energy by synchrotron emission and inverse Compton emission. 
The latter process involves the upscattering of photons from the hot stellar companion to gamma-ray energies.
The exact origin of the flux variability is much harder to identify, as it is very likely due to the combination of several physical processes;
the most relevant of which are discussed in the following.

The acceleration of charged particles in binary system is generally attributed to first order Fermi acceleration in  strong shocks formed
between the stellar wind and the wind of the pulsar (assuming that the compact object in HESS J0632+057 is a fast rotating neutron star).
The shock conditions are strongly dependent on the geometry of the system, and vary significantly during one orbital period.
The orbital movement and the  Coriolis forces involved can lead to the formation of several shocks in regions of varying gas and photon densities \cite{BoschRamon-2012, Zabalza-2012}.
Further away from the shocked regions, acceleration of charged particles by occurring velocity shears of the relativistic outflows have been predicted \cite{Rieger-2006}.
High accretion rates during the passage of the compact object through the stellar disk or close to the massive star can lead to a quenching of the pulsar wind and might be responsible for the sharp dip in the X-ray emission at phases $\approx 0.45$. 
Note that the extension, density and orientation of the disk of the Be star in  HESS J0632+057 is mostly unknown,
although recent optical dispersion measurements indicate that the disk of the Be star is likely as large as the binary orbit \cite{Moritani-2015}.
This setting is very similar to the flip-flop scenario suggested for the gamma-ray binary LS I +61 303 \cite{Torres-2012}, where periods in which the system is in the  rotationally powered regime alternate with the propeller regime.
In addition, absorption of high-energy gamma rays due to pair production is expected to be most important around periastron and inferior conjunction, where HESS J0632+057 is in a low flux state (see e.g.~\cite{Dubus-2010}).

In summary, new and updated observations of the gamma-ray binary HESS J0632+057 by VERITAS and {\em Swift} XRT provide significantly better measurements of the complex situation in this gamma-ray binary.
VERITAS will continue to observe HESS J0632+057 as part of the observatories' long-term observing plan over the coming years.

\subsubsection*{Acknowledgments}

This research is supported by grants from the U.S. Department of Energy Office of Science, the U.S. National Science Foundation and the Smithsonian Institution, and by NSERC in Canada. 
G.M.~acknowledges support through the Young Investigators Program of the Helmholtz Association. 
We acknowledge the excellent work of the technical support staff at the Fred Lawrence Whipple Observatory and at the collaborating institutions in the construction and operation of the instrument.
The VERITAS Collaboration is grateful to Trevor Weekes for his seminal contributions and leadership in the field of VHE gamma-ray astrophysics, which made this study possible. 

This work made use of data supplied by the UK Swift Science Data Centre at the University of Leicester.


\begin{thebibliography}{99}
\bibitem{Dubus-2013} Dubus, G. 2013, Astron Astrophys Rev 21, 64
\bibitem{Aragona-2010} Aragona, C., McSwain, M. V., \& De Becker, M. 2010, ApJ, 724, 306
\bibitem{Aharonian-2007} Aharonian, F. et al (H.E.S.S. collaboration) 2007, A\&A, 469, L1
\bibitem{Acciaro-2009} Acciari, V.A. et al (VERITAS collaboration) 2009, ApJ 498, L94
\bibitem{Bongiorno-2011} Bongiorno, S.D. et al., ApJ 2011, 737, L11
\bibitem{Aliu-2014} Aliu, E. et al.\ (VERITAS \& H.E.S.S. Collaborations 2014, ApJ, 780, 168
\bibitem{Aleksic-2012} Aleksi\'{c}, J. et al.\  (MAGIC Collaboration) 2012, ApJ, 754, L10
\bibitem{Casares-2012} Casares, J. et al. \ 2012, MNRAS 421, 1103
\bibitem{Caliandro-2013} Caliandro, G.A. et al. \ 2013, MNRAS 436, 740
\bibitem{Staszak:2015} Staszak, D. et al.\ (VERITAS Collaboration) 2015, 34th ICRC in The Hague
\bibitem{Park:2015} Park, N. et al.\ (VERITAS Collaboration) 2015, 34th ICRC in The Hague
\bibitem{WWW} http://veritas.sao.arizona.edu/specifications
\bibitem{Swift} {Evans}, P.~A. et al. 2007, A\&A 469, 379
\bibitem{Alexander-1997} Alexander, T. 1997, in Astronomical Time Series, ed. D. Maoz, A. Sternberg, \& E. M. Leibowitz (Dordrecht: Kluwer), 163
\bibitem{Stellingwerf-1978} Stellingwerf, R.F. 1987, ApJ, 224, 953
\bibitem{BoschRamon-2012} Bosch-Ramon, V. et al. \ 2012, A\&A, 544, A59
\bibitem{Zabalza-2012} Zabalza, V. et al. \ 2013, A\&A, 551, A17 
\bibitem{Rieger-2006} Rieger, F. \& Duffy, P. \ 2006, ApJ 652, 1044
\bibitem{Moritani-2015} Moritani, Y. et al. \ 2015, ApJ 804, L32
\bibitem{Torres-2012} Torres, D. et al \ 2012, ApJ 744, 106
\bibitem{Dubus-2010} Dubus, G. \ 2008, A\&A, 477, 691


\end{thebibliography}
\end{document}